\documentclass[10pt,apjl]{emulateapj}

\usepackage{apjfonts}

\newcommand{\teff}{$T_{\rm eff}$}

\shorttitle{CNO abundances in NGC 1851}
\shortauthors{Yong et al.\ }

\begin{document}

\title{A large C+N+O abundance spread in giant stars of the globular 
cluster NGC 1851\altaffilmark{1}}

\author{
David Yong\altaffilmark{2},
Frank Grundahl\altaffilmark{3},
Francesca D'Antona\altaffilmark{4},
Amanda I.\ Karakas\altaffilmark{2}, \\
John C.\ Lattanzio\altaffilmark{5}, and 
John E.\ Norris\altaffilmark{2}
}

\altaffiltext{1}{Based on 
observations made with the Magellan Clay Telescope at Las Campanas 
Observatory.}

\altaffiltext{2}{Research School of Astronomy and Astrophysics, 
Australian National University, Weston, ACT 2611, Australia; 
yong@mso.anu.edu.au, akarakas@mso.anu.edu.au, jen@mso.anu.edu.au}
\altaffiltext{3}{Department of Physics and Astronomy, Aarhus University, 
Denmark; fgj@phys.au.dk}
\altaffiltext{4}{INAF-Osservatorio Astronomico di Roma, via Frascati 33, 
00040 Monteporzio, Italy; dantona@mporzio.astro.it}
\altaffiltext{5}{Centre for Stellar and Planetary Astrophysics, 
School of Mathematical Sciences, Monash University, Victoria 3800, Australia; 
John.Lattanzio@sci.monash.edu.au}

\begin{abstract}

Abundances of C, N, and O are determined in four bright red giants 
that span the known abundance range for light (Na and Al) and 
$s$-process (Zr and La) elements in the globular cluster NGC 1851. 
The abundance sum C+N+O exhibits a range of 0.6 dex, a factor of 4, in 
contrast to other clusters in which no significant C+N+O spread is found. 
Such an abundance range offers support for the \citet{cassisi08} 
scenario in which the double subgiant branch populations 
are coeval but with different mixtures of C+N+O abundances. 
Further, the Na, Al, Zr, and La abundances are correlated 
with C+N+O, and therefore, NGC 1851 is the first cluster to provide 
strong support for the scenario in which AGB stars 
are responsible for the globular cluster light element abundance variations. 

\end{abstract}


\section{Introduction}
\label{sec:intro}

Globular clusters have long been regarded as simple stellar populations 
which may be described by a single age, helium abundance ($Y$), 
metallicity ($Z$), and initial mass function \citep{rb86}. For many years 
these simple stellar populations have played a prominent role in astronomy 
by providing a lower limit to the age of the Universe and for testing the 
predictions of stellar evolution and stellar nucleosynthesis (e.g., see review 
by \citealt{gratton04}). 

However, 
a revolution in the field of globular cluster research is underway, 
prompted by the recent 
discoveries of complex structure on the main 
sequence, subgiant branch (SGB), red giant branch (RGB), 
and/or horizontal branch 
(HB) within some Galactic and extra-Galactic globular clusters 
indicating that these simple stellar populations in fact contain 
discrete, multiple populations 
(e.g., \citealt{bedin04,dantona05,piotto05,sollima05,mackey08}). 
At present, these multiple populations can be best explained in terms 
of distinct ages and/or compositions, although the sequence of events 
leading to their formation remains largely unexplained \citep{renzini08}. 
In some cases, extremely 
large helium abundances, up to $Y$ = 0.40, are required to explain 
the various populations (e.g., \citealt{dantona04,norris04}). While   
theoretical models have struggled to account for such high 
abundances (e.g., \citealt{karakas06,bekki07}), some successful 
efforts are beginning to appear (e.g., \citealt{dercole08,pumo08}). 
The intrigue surrounding the multiple populations in globular clusters has 
been enhanced by the speculation that clusters with extended HBs, 
generally the most massive clusters, 
are the remnants of dwarf galaxies from which the halo of the Galaxy was 
built \citep{lee07}. 

The globular cluster NGC 1851 has a bimodal HB and 
displays 
multiple SGBs \citep[hereafter M08]{milone08}, 
but no evidence for 
multiple main sequences. 
Based on the width of the main sequence and 
RGB, M08 suggested that the maximum possible 
abundance variation is $\Delta$[Fe/H] = 0.1 dex or $\Delta Y$ = 0.026. 
From an analysis of eight bright RGB stars using high resolution 
spectroscopy, \citet[hereafter YG08]{yg08} 
showed that the dispersion in [Fe/H] was less than 0.1 dex, 
and that NGC 1851 displays the 
usual star-to-star abundance variations for the light elements O, Na, 
and Al; the exact mechanism for these 
abundance patterns, seen in all globular clusters, 
remains unknown (e.g., \citealt{gratton04}). 
However, YG08 also showed that NGC 1851 harbors a star-to-star abundance 
variation for the $s$-process elements Zr and La and that the abundances 
of these elements were correlated with the light element abundances. 
Furthermore, within the small sample 
there was a hint that the abundances of the $s$-process elements was bimodal, 
which suggests that the RGB may consist of two populations with 
distinct chemical compositions. 
Therefore, as well as showing unusual photometric properties, NGC 1851 
also displays peculiar chemical abundance patterns not seen in other clusters. 

Several explanations for the 
double SGB in NGC 1851 have been offered. 
If age is the sole parameter, 
M08 suggested a 1 Gyr difference 
with the fainter subgiant branch (fSGB) population 
being older. 
M08 also considered a combination of increasing [Fe/H] by 0.2 dex and 
helium from $Y$ = 0.247 to 0.30, 
though this 
possibility was excluded by the magnitudes of the blue HB stars. 
An alternative explanation was proposed by \citet{cassisi08} 
in which the two SGB populations 
are coeval, but with the fSGB population having a 
total C+N+O abundance increased by a factor of 2. 
In this scenario, the two 
populations may have comparable He abundance 
(at most $\Delta Y$ = 0.032), in contrast to 
the large He variations inferred in 
other clusters which display multiple populations. We have obtained 
spectra with the goal of measuring the abundance sum C+N+O to test the 
\citet{cassisi08} scenario. 

\section{Observations and abundance analysis}
\label{sec:data}

\begin{figure}
\epsscale{1.1}
\plotone{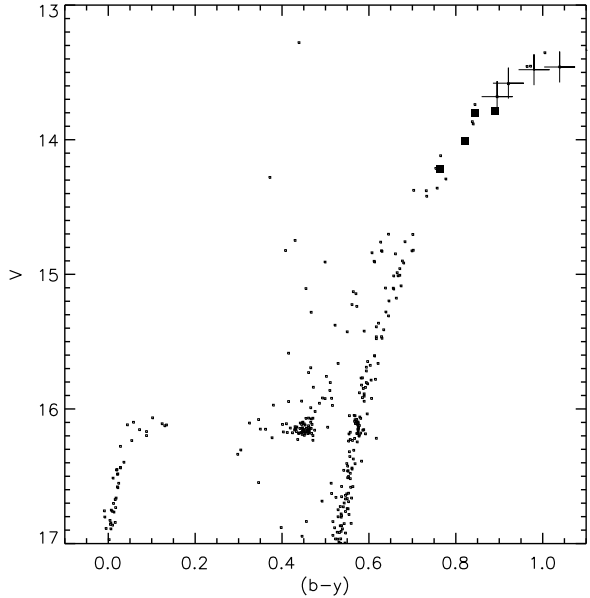}
\caption{The color-magnitude diagram for NGC 1851 using the 
\citet{grundahl99} photometry. The large plus symbols represent 
the bright giants in this study. The filled squares represent the 
additional four stars in YG08. \label{fig:cmd}}
\end{figure}

We selected four bright RGB 
stars from YG08, two Na-, Al-, Zr-, and La-normal targets 
as well as two Na-, Al-, Zr-, and La-rich objects. 
In Figure \ref{fig:cmd} we show the locations of our targets 
in a color-magnitude diagram. 
If the double SGB populations are due to distinct 
chemical compositions, and if these chemical compositions are also 
present on the RGB, 
then our small, but carefully selected, sample should allow us to 
identify any C+N+O abundance variation, if present. 

The stars were 
observed with the high resolution spectrograph MIKE \citep{mike} on the 
Magellan Clay Telescope on 2007 December 23. We used the 0.7\arcsec\ slit 
which provided a spectral resolution of $R$ $\simeq$ 37,000 in the blue CCD 
(3350-5000\AA) and $R$ $\simeq$ 30,000 in the red CCD (4900-9500\AA) as 
measured from the ThAr lines. Our total 
exposure times per star ranged from 40 to 80 minutes resulting in 
signal-to-noise ratios (S/N) per four pixel resolution element of 
70-100 at 4320\AA\ and 210-260 at 6600\AA. The data were reduced using the 
same procedure described by \citet{yong06}. 

Although the new data are of lower spectral resolution 
($R$ $\simeq$ 30,000 vs.\ $R$ $\simeq$ 55,000), 
the S/N and wavelength coverage are superior to the 
spectra analyzed by YG08. Using these new spectra, we rederived the 
stellar parameters using the same tools and techniques described in YG08 
and found that 
the revised stellar parameters are in good agreement with the 
stellar parameters previously obtained. The new dispersion 
in Fe abundances for these four stars is $\sigma$[Fe/H] = 0.03 dex 
compared with 0.11 dex for the YG08 analysis. This decrease is 
due to the increased wavelength coverage and S/N which allows us to more 
accurately measure the equivalent widths of a larger number of Fe lines. 

\begin{figure}[h!]
\epsscale{1.1}
\plotone{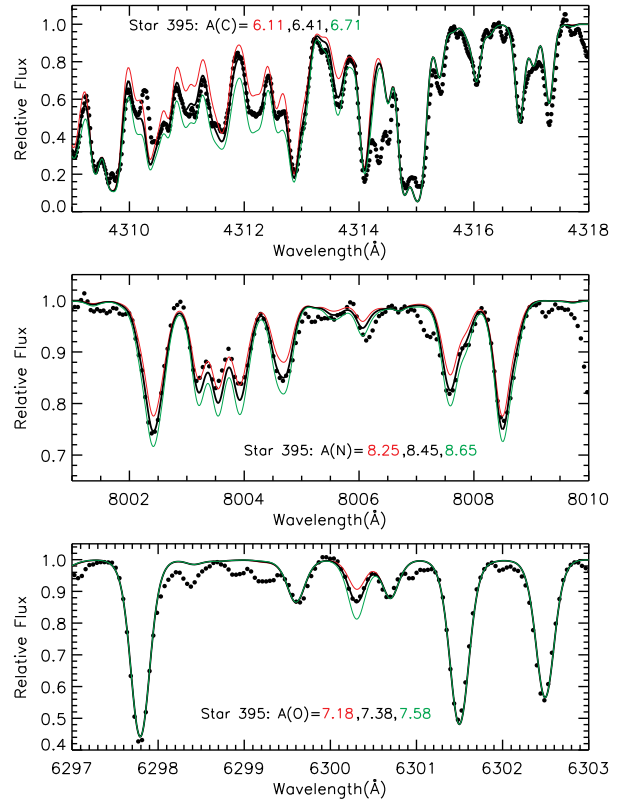}
\caption{Observed spectra (circles) and synthetic spectra (lines) for 
C (upper), N (middle), and O (lower) for Star 395. 
The synthetic spectra show the best fit (thick black line) and unsatisfactory 
fits (thin red and green lines) $A$(C) $\pm$ 0.30 dex, $A$(N) $\pm$ 0.20 dex, 
and $A$(O) $\pm$ 0.20 dex. \label{fig:cnospec1}}
\end{figure}

\begin{figure}[h!]
\epsscale{1.1}
\plotone{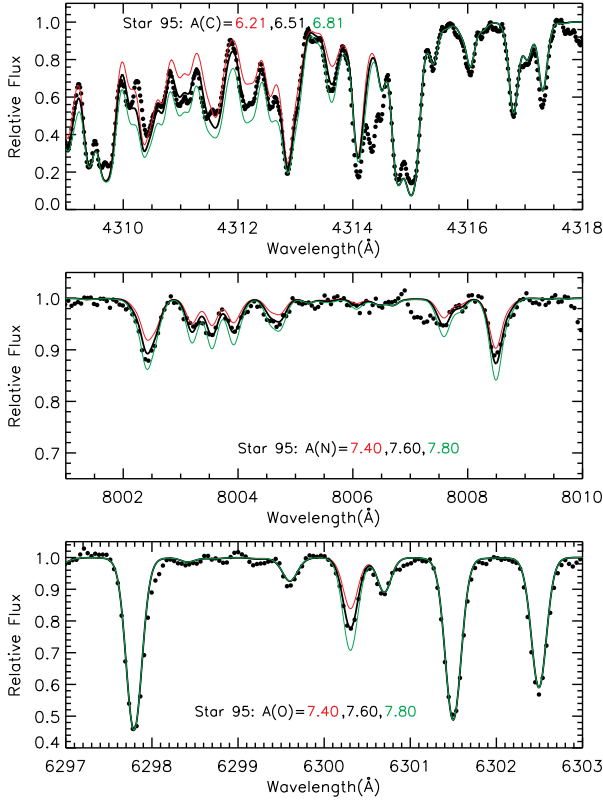}
\caption{Same as Figure \ref{fig:cnospec1} but for Star 95.
\label{fig:cnospec2}}
\end{figure}

The abundances of C and N were derived by comparing 
observed spectra with synthetic spectra generated 
using the LTE line analysis and spectrum synthesis program MOOG \citep{moog}. 
For C, we analyzed lines from 
the $(0,0)$ and $(1,1)$ bands of the $A-X$ electronic 
transition of the CH molecule near 4310\AA\ using the line list from 
\citet{ch}. For N, we analyzed lines from 
the $(2,0)$ band of the $A-X$ electronic 
transition of the CN molecule near 8000\AA\ using the line list from 
\citet{reddy02} which is based on the \citet{cn} list. 
In both cases, we adjusted the abundances of C or N until the 
synthetic spectra matched the observed spectra 
(see Figures \ref{fig:cnospec1} and \ref{fig:cnospec2}). 
We rederived the O abundances again from the 6300\AA\ [OI] line 
using the new spectra, the revised C 
and N abundances, and the revised stellar parameters. 
Since the abundances of C, N, and O are coupled, we 
iterated until self-consistent abundances were obtained. 
We also updated the abundances for Na, Al, Zr and La 
using the new spectra and revised stellar parameters 
(a complete abundance analysis of our data will be 
presented elsewhere). We present the stellar parameters and abundances 
in Table \ref{tab:abund}. 

\begin{deluxetable*}{cccccccccccccc}
\tablecolumns{14} 
\tablewidth{0pc} 
\tablecaption{Stellar parameters and abundances\label{tab:abund}} 
\tablehead{ 
\colhead{Star\tablenotemark{a}} & 
\colhead{$V$} & 
\colhead{\teff~(K)} & 
\colhead{$\log g$ (cgs)} & 
\colhead{$v_t$ (km s$^{-1}$)} & 
\colhead{$A$(Fe)} &
\colhead{$A$(C)} &
\colhead{$A$(N)} &
\colhead{$A$(O)} &
\colhead{$A$(CNO)} &
\colhead{$A$(Na)} &
\colhead{$A$(Al)} &
\colhead{$A$(Zr)} &
\colhead{$A$(La)} 
}
\startdata 
003 & 13.68 & 4125 & 0.7 & 1.86 & 6.28 & 6.46 & 7.60 & 7.75 & 8.00 & 5.25 & 5.63 & 1.51 & 0.16 \\
095 & 13.58 & 4125 & 0.7 & 1.92 & 6.32 & 6.51 & 7.60 & 7.60 & 7.92 & 5.25 & 5.69 & 1.64 & 0.22 \\
329 & 13.46 & 3925 & 0.0 & 1.92 & 6.25 & 6.81 & 8.00 & 7.88 & 8.26 & 5.50 & 5.79 & 1.87 & 0.42 \\
395 & 13.48 & 4025 & 0.5 & 2.01 & 6.29 & 6.41 & 8.45 & 7.38 & 8.49 & 5.74 & 6.04 & 1.85 & 0.53 
\enddata 
\tablenotetext{a}{Star names taken from \citet{stetson81}.}
\end{deluxetable*} 

\section{Results }
\label{sec:results}

As mentioned, our sample was selected to span the full range of the 
known star-to-star abundance variation of Na, Al, Zr, and La. 
Our new analysis confirms the large abundance spread for these
elements as well as the correlations between the 
light (Na and Al) and heavy (Zr and La) elements established by 
YG08. The range in Na, Al, Zr, and La abundances 
far exceeds the measurement uncertainties of 0.06, 0.05, 0.12, and 0.08 
dex respectively, though we note that the abundance amplitudes for 
these elements are slightly smaller than those reported by YG08. 

Adopting a solar abundance 
$\log~\epsilon$(Fe) = 7.50 \citep{grevesse98}, 
we find a mean cluster abundance 
of [Fe/H] = $-$1.22 ($\sigma$ = 0.03). The revised cluster Fe 
abundance is in good agreement with the YG08 value and previous 
estimates in the literature, but the 
abundance dispersion is considerably smaller for the 
reasons noted above. The abundance range for Fe may be entirely 
attributable to observational uncertainties (0.07 dex). 

For C and N, we find a large range in abundances. The 0.40 dex 
range in $\log~\epsilon$(C) and the 0.85 dex range in $\log~\epsilon$(N) 
greatly exceed the measurement uncertainties 
(0.04 dex and 0.11 dex respectively). 
The abundance amplitude for N, $\Delta\log~\epsilon$(N) = 0.85, is smaller 
than that seen in other clusters, e.g., RGB bump stars in NGC 6752 and main 
sequence stars in M13 show 2.0 dex amplitudes \citep{briley02,6752nh}. 
This may be due to the small sample 
and/or advanced evolutionary status where extensive CN-cycling may 
reduce the amplitude of the N abundance dispersion. 
Curiously, we do not detect any obvious anticorrelation between C and N 
as seen in every well studied Galactic globular cluster. We 
identify correlations between the abundances of N and Na as well as N 
and Al, but the corresponding anticorrelations between C and Na and 
C and Al are not obvious. Nevertheless, we find correlations between 
the abundances of N and Zr as well as N and La as expected 
given the known behavior of N and Al and the 
correlations between Al and Zr and Al and La found YG08. 

Our first significant result 
is that the abundance sum 
C+N+O shows a large variation. Within our small but biased sample, 
the amplitude of the abundance variation is $\Delta$$A$(C+N+O) = 0.57 dex. 
The 0.57 dex spread far exceeds the estimated uncertainty of 0.14 dex. 
Bright giants in M4 and NGC 6712, 
globular clusters of comparable metallicity to NGC 1851, 
show C+N+O abundance amplitudes of 0.22 dex and 0.35 dex respectively 
\citep{smith05,6712}. Unevolved stars in NGC 6397, 
NGC 6752 and 47 Tuc show C+N+O abundances that 
may be regarded as constant within the $\sim$0.3 dex 
uncertainties \citep{carretta05}. 

\begin{figure}
\epsscale{1.1}
\plotone{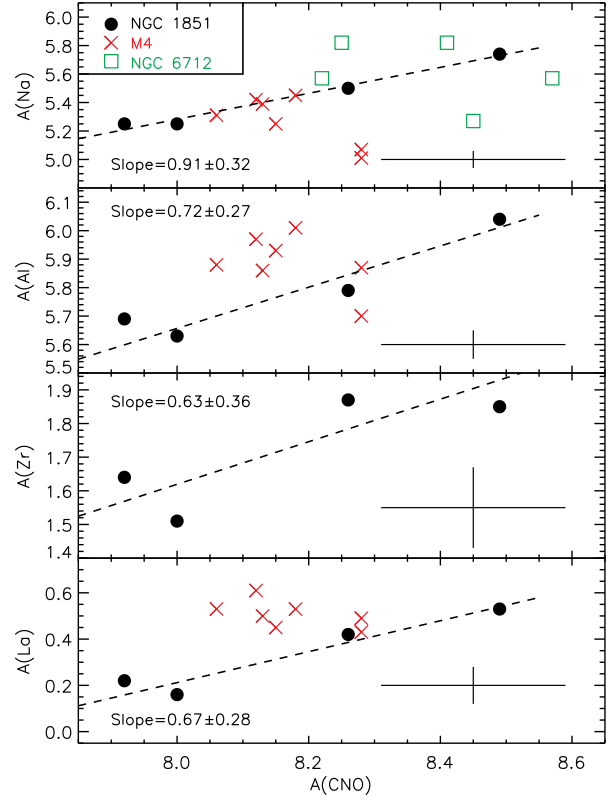}
\caption{The abundances of Na, Al, Zr, and La vs.\ the abundance sum C+N+O. 
A representative error bar and a linear least squares fit 
to the data (including formal slope and error) are shown. Data for 
M4 \citep{smith05} and NGC 6712 \citep{6712} are included for comparison. 
\label{fig:cno}}
\end{figure}

Our second main result 
is that the abundances of 
Na, Al, Zr, and La show a positive correlation with the abundance sum C+N+O 
(see Figure \ref{fig:cno}). 
In this figure, we overplot results for M4 and NGC 6712 and find that for 
the available data, there are no positive correlations between $A$(X) and C+N+O 
in these other clusters. Finally, our small sample does not exclude 
the possibility that the C+N+O abundance may be bimodal within this 
cluster. 

\section{Discussion}
\label{sec:discussion}

We seek to test the \citet{cassisi08} prediction in which 
the double SGB in NGC 1851 
is due to two coeval populations whose chemical compositions 
for C+N+O differ by a factor of two. 
Note that we have not directly observed stars on the SGB. 
Instead, we have observed a sample of bright RGB stars in which we have 
found a large spread in the C+N+O abundances. If the SGBs  
comprise populations with different C+N+O abundances, then we 
assume that these populations with distinct compositions are also 
present on the main sequence and RGB. Str{\" om}gren photometry 
obtained by \citet{grundahl99} 
reveals that the bimodality seen on the SGB can also be 
found on the RGB. Therefore, 
the large spread in C+N+O abundances in RGB stars offers 
support, albeit not definitive, for the \citet{cassisi08} scenario. 
We tentatively confirm that the double SGB  
in NGC 1851 is due to two coeval populations with C+N+O abundances 
differing by a factor of two. 

Based on the relative numbers of stars, YG08 speculated that the 
brighter subgiant branch (bSGB) stars were related to the RGB stars with 
``normal'' Zr and La abundances and that these stars populate the 
red HB. YG08 also suggested that the fSGB stars 
are related to the RGB stars with Zr and La enhancements and that these 
stars populate the blue HB. 
If the fSGB is older than the bSGB as M08 suggest, then the YG08 
speculation would require the Zr-rich,
La-rich stars to be older than the Zr-normal, La-normal stars. 
In this scenario, the sequence of events leading to cluster formation 
would be complicated and require the AGB (or other) stars to 
preferentially pollute the older Zr-rich La-rich population 
but without polluting the gas from which the 
younger Zr-normal La-normal population formed. 
However, if the C+N+O abundance of the $s$-process rich
stars is larger than that of the $s$-process normal ones, 
the fSGB may be coeval
to the bSGB, or even a bit younger than the bSGB, as shown in Cassisi
et al. (2008). Due to the observational errors, and to the fact that
the age difference we are looking for does not exceed some 100Myr, it
may be very difficult or impossible to constrain the precise mass
range of the progenitors of the $s$-process rich stars, but a coherent picture 
of the chemical evolution history is emerging.

The \citet{cassisi08} scenario 
has important implications for other clusters which 
display multiple SGBs. \citet{villanova07} studied the 
SGB region of $\omega$ Centauri, the most massive globular cluster 
which is suspected of harboring up to five populations with discrete 
ages and/or compositions. From the large range in the magnitudes 
of the SGB populations, \citet{villanova07} inferred a large 
range in ages. They found that there was an extremely old, but metal-rich 
population which greatly complicates any formation scenario. 
Based on our NGC 1851 results, the inferred ages of SGB 
stars in $\omega$ Cen requires detailed knowledge of the relative 
abundances of C+N+O. We speculate that the old metal-rich population 
(Villanova et al.'s ``SGB group D'') has a large excess of 
[(C+N+O)/Fe] relative to the more metal-poor populations and therefore a 
younger relative age than currently inferred. Such compositions 
would be consistent with our understanding of the composition of the ejecta 
from AGB stars \citep{karakas07} 
and the prominent role of AGB stars in the chemical evolution 
of $\omega$ Cen \citep{norris95}. 

AGB stars have long been suspected as being responsible, in part or in 
whole, for the star-to-star abundance variation for the light elements 
(C, N, O, F, Na, Mg, and Al) found in every well studied Galactic 
globular cluster, including NGC 1851. However, various theoretical yields 
and chemical evolution models involving AGB stars have yet to provide 
a satisfactory solution (e.g., \citealt{fenner04}). AGB stars may be 
expected to produce a substantial increase in the C+N+O abundance as they 
enhance Na and Al and deplete O and Mg. AGB stars should also produce 
$s$-process elements, albeit via different processes than for 
the O, Na, Mg, and Al nucleosynthesis. If AGB stars are responsible 
for the light element abundance variations in globular clusters, 
we may expect 
a large range in C+N+O abundances along 
with correlations between (1) C+N+O and Na, 
(2) C+N+O and Al, and (3) C+N+O and $s$-process elements. 
On the other hand, these proposed abundance trends are dependent on the 
number of third dredge up episodes 
and therefore 
on the adopted convection efficiency 
(e.g., \citealt{ventura05a,renzini08}) 
and mass loss 
(e.g., \citealt{ventura05b}). 
Further 
uncertainty in the yields comes from the uncertainties in the nuclear 
reaction rates 
(e.g., \citealt{lugaro04,ventura06,karakas06mg,izzard07}).  
Although the brightest AGB stars in the LMC and SMC can become 
C-rich, it is 
possible that nucleosynthesis in the most massive AGBs is scarcely 
affected by the third dredge up, with only a minor increase in the total 
C+N+O abundance of the second generation stars 
\citep{dantona08}. 
Nevertheless, the absence of 
the signatures listed above has forced astronomers to 
consider
alternative sources such as massive stars 
(e.g., \citealt{decressin07}). 

The correlations between C+N+O and Na, Al, Zr, and La presented here 
offer strong support for the scenario in which AGB stars are responsible 
for the light element abundance variations. Given the peculiar photometric 
and chemical properties of NGC 1851, it is surprising that this cluster 
offers the strongest support for the globular cluster AGB pollution scenario. 
It would be ironic if NGC 1851 is the only globular cluster for which 
the AGB pollution scenario is applicable. An alternative interpretation 
is that the AGB pollution scenario is applicable to all globular clusters. 
For NGC 1851, the masses of the polluting AGB stars was lower such that 
the products of third dredge-up, increased C+N+O and $s$-process elements, 
are clearly evident. 

We note that abundance spreads in $A$(CNO) are likely present 
in $\omega$ Centauri \citep{norris95}, although they were  
concerned with possible problems with the N measurements. 
Indeed, NGC 1851 bears a 
striking resemblance to $\omega$ Centauri. Both clusters possess 
multiple SGBs, large abundance spreads for 
$s$-process elements and likely spreads in $A$(CNO), 
and a scatter in the Str{\" om}gren $m_1$ index, 
traditionally used as a metallicity indicator \citep{grundahl99}. 
We reiterate the comments by YG08 that NGC 1851 is a ``bridge'' between 
$\omega$ Centauri (variation of all elements) and NGC 6752-like clusters 
(constant Fe and C+N+O but large variations of light elements C-Al). 

Finally, recent work by \citet{salaris08} also suggests a 
relationship between the SGB and HB 
populations in NGC 1851. Specifically, the bSGB stars 
evolve onto the red HB while the fSGB stars populate 
the red and blue HB. If this is correct, we would therefore expect 
stars on the red HB to be primarily Zr-normal and La-normal. NGC 1851 
may provide critical information on HB morphology 
and the second parameter effect and 
additional studies of this cluster are of great interest. 

\acknowledgments

DY thanks Bertrand Plez for providing the CH line list and 
Bacham Reddy for providing the CN line list. 
We thank the anonymous referee for helpful comments. 
FG acknowledges financial support from the Carlsberg Foundation, the
Danish AsteroSeismology Centre at Aarhus University and support from the 
Danish National Research Council to the project: "Stars: Cental engines of 
the evolution of the  Universe." 
FD acknowledges partial support from  the italian PRIN MIUR 2007
'Multiple stellar populations in globular clusters: census,
characterization and
origin'. 
We acknowledge support from the Australian Research Council's 
Discovery Projects funding scheme 
under grants DP0663562 (JEN) and DP0664105 (AIK).

\end{document}